\documentclass[twocolumn,conference]{IEEEtran}
\usepackage[T1]{fontenc}
\usepackage[latin9]{inputenc}
\usepackage{color}
\usepackage{array}
\usepackage{booktabs}
\usepackage{multirow}
\usepackage{amstext}
\usepackage{graphicx}

\usepackage{lipsum,filecontents}

\makeatletter

\providecommand{\tabularnewline}{\\}

\usepackage[caption=false,font=footnotesize]{subfig}
\usepackage{cite}

\makeatother

\begin{document}
\IEEEoverridecommandlockouts
\IEEEpubid{\makebox[\columnwidth]{979-8-3503-9958-5/22/\$31.00~\copyright2022 IEEE\hfill} \hspace{\columnsep}\makebox[\columnwidth]{ }}
\title{Dynamic Programmable Wireless Environment with UAV-mounted Static Metasurfaces}% \\ {\large Short Paper}}
\author{ Prodromos-Vasileios Mekikis\IEEEauthorrefmark{1}, Dimitrios Tyrovolas\IEEEauthorrefmark{1}, 
Sotiris Tegos\IEEEauthorrefmark{1}, Alexandros Papadopoulos\IEEEauthorrefmark{2},\\
Alexandros Pitilakis\IEEEauthorrefmark{1}, Sotiris Ioannidis\IEEEauthorrefmark{3}\IEEEauthorrefmark{4},
Ageliki Tsioliaridou\IEEEauthorrefmark{3}, Panagiotis Diamantoulakis\IEEEauthorrefmark{1},\\
Nikolaos Kantartzis\IEEEauthorrefmark{1}, George Karagiannidis\IEEEauthorrefmark{1},
Christos Liaskos\IEEEauthorrefmark{2}\IEEEauthorrefmark{3}\\
{\small\IEEEauthorrefmark{1}Aristotle University of Thessaloniki,
emails: {[}vmekikis, tyrovolas, tegosoti, alexpiti, padiaman, kant,
geokarag{]}@ece.auth.gr}\\
{\small\IEEEauthorrefmark{2}University of Ioannina, emails: {[}a.papadopoulos,
cliaskos{]}@uoi.gr}\\
{\small\IEEEauthorrefmark{3}Foundation for Research and Technology
- Hellas (FORTH), emails: {[}sotiris, atsiolia{]}@ics.forth.gr}\\
{\small\IEEEauthorrefmark{4}Technical University of Chania, Crete}}
\maketitle
\begin{abstract}
Reconfigurable intelligent surfaces (RISs) are artificial planar structures able to offer a unique way of manipulating propagated wireless signals. Commonly composed of a number of reconfigurable passive cell components and basic electronic circuits, RISs can almost freely perform a set of wave modification functionalities, in order to realize programmable wireless environments (PWEs). However, a more energy-efficient way to realize a PWE is through dynamically relocating static metasurfaces that perform a unique functionality. In this paper, we employ a UAV swarm to dynamically deploy a set of low-cost passive metasurfaces that are able to perform only one electromagnetic functionality, but with the benefit of requiring no power. Specifically, the UAV-mounted static metasurfaces are carefully positioned across the sky to create cascaded channels for improved user service and security hardening. The performance evaluation results, based on ray tracing, demonstrate the potential of the proposed method.
\end{abstract}


\section{Introduction}

Recently, a wireless communication paradigm has emerged, which strives to constitute the wireless propagation phenomenon into a software-defined and deterministic process~\cite{liaskos2019network}. This direction, denoted as programmable wireless environments (PWEs), is enabled via coating planar objects in a space with metasurfaces, more recently
known as reconfigurable intelligent surfaces (RISs). RISs constitute a merge of metasurfaces with well-defined networking and programming interfaces. Employing the Huygens principle, RIS can manipulate impinging waves by altering their power, direction, polarization, and phase~\cite{MSSurveyAllFunctionsAndTypes}. Thus, RIS can be dynamically orchestrated to realize custom end-to-end propagation routes, e.g., to avoid unintended users, increase the quality of the wireless channels,  or even realize futuristic applications such as extended reality~\cite{liaskos2019network, trung, XR}. 

PWEs are presently combined with flexible unmanned aerial vehicles (UAVs) which carry RIS units to carefully selected places and configure their electromagnetic behavior dynamically~\cite{yang2020performance}, \cite{Harq},\cite{alouini}. While promising, this approach burdens with additional cost and power consumption an already high-cost and energy-restricted UAV system. In this work, we explore a direction where only inexpensive, static metasurface units are required to dynamically create a PWE. Specifically, inspired by late trends in UAV-based delivery of goods~\cite{san2018uav}, we consider a set of UAVs, where each one can dynamically position a metasurface, which can only serve a specific wave manipulation functionality, such as steering an impinging wave to a specific direction of departure. Therefore, by following an orchestration process, the UAVs can fine-tune these static metasurface units properly to form cascading channels~\cite{tyrovolas2022performance} and, thus, improve the quality-of-service (QoS) of the network users in terms of connectivity and security. Hence, compared to synergetic UAV-RIS networks, the proposed approach has the benefit of low-cost and simplicity, as the UAV-orchestrated metasurfaces can transform a wireless propagation environment into a dynamic PWE.

The remainder of this paper is as follows. Section~\textcolor{black}{\ref{sec:The-approach} details the proposed scheme. Evaluation follows in Section\ \ref{sec:Evaluation} and the conclusion in Section\ \ref{sec:Conclusion}.}

\section{UAV-Deployable PWE with static metasurfaces \protect\label{sec:The-approach}}

\begin{figure*}[!t]
\centering{}\textcolor{black}{\includegraphics[clip,width=0.8\textwidth]{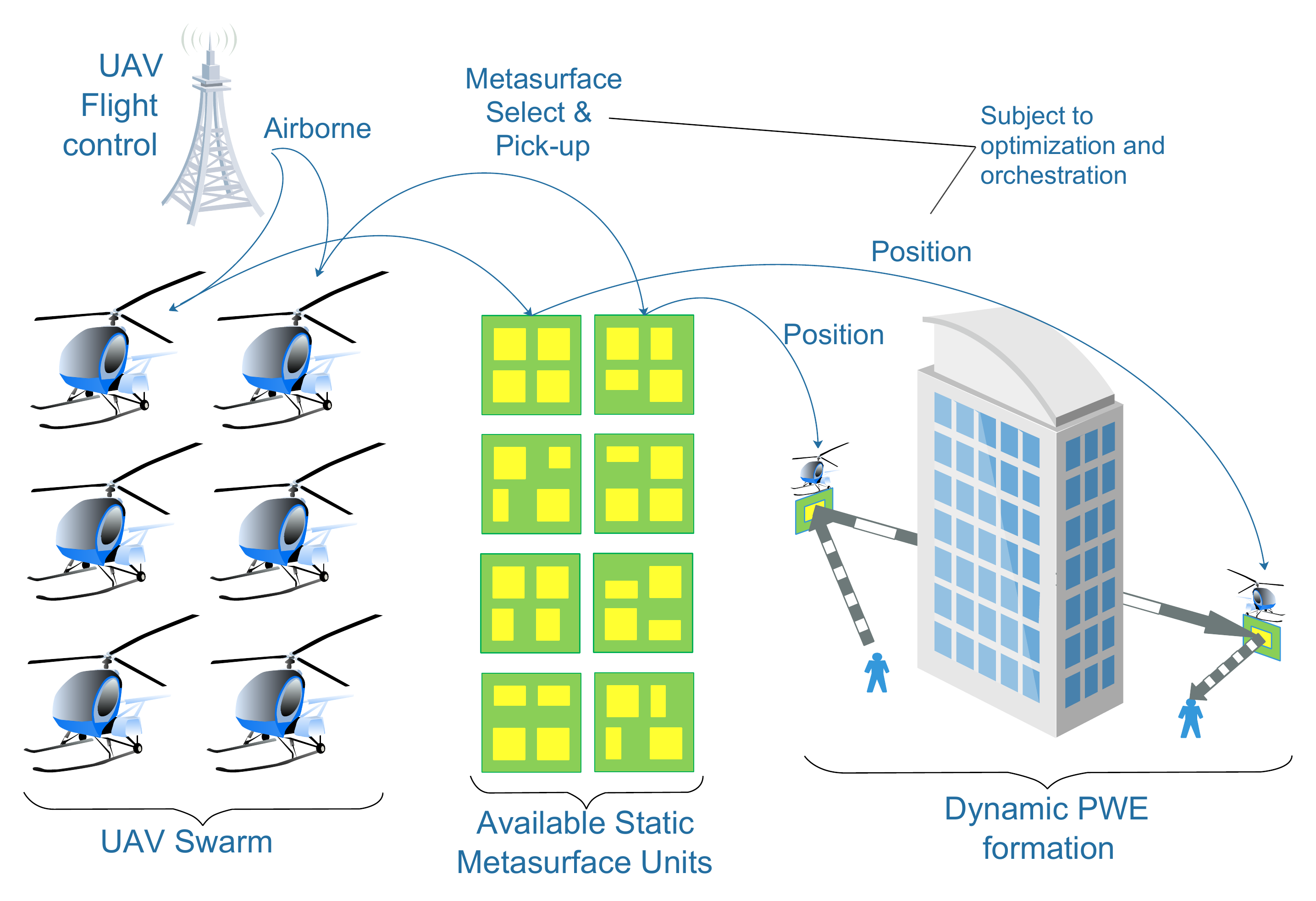}
\caption{\protect\label{fig:structure}SDN schematic displaying the system
model and workflow abstraction.}
}
\end{figure*}
Fig.~\ref{fig:structure} represents a schematic of the proposed
system architecture. We consider an environment hosting wireless users,
with multiple blockages present. In this setting, a UAV swarm
is licensed to operate under the command of a UAV flight control
system. Each UAV in the swarm can be dynamically equipped with
one static metasurface, available at the premises of the UAV ground station. Each metasurface can serve only one static wave manipulation functionality,
such as steering from and to very specific wave directions of arrival
and departure. 

The UAV flight control is assumed to be delegated a communication
assistance task by a third-party communication system, which monitors
its users and deduces the connectivity requirements. Subsequently, it invokes
the UAV flight control to dynamically deploy a PWE in the premise
of the users in need of assistance or communication augmentation.
The UAV flight control then executes a UAV-driven PWE orchestration
algorithm and deduces the static metasurfaces to be deployed at the user premises,
in a way that end-to-end cascaded air links and channels are formed.
Following the orchestration planning phase, the UAV control instructs
a set of UAVs in the swarm to pick up the selected RIS units and
move to positions designated by the orchestration process. Once there,
the channel is deployed automatically, without further deliberations.

We proceed to detail the specifications and workflow of each component of the proposed architecture, i.e., the UAV flight control and orchestration system, the static metasurface units, and the dynamic PWE orchestration and configuration process.

\subsection{UAV flight control and orchestration system \protect\label{subsec:Compressive-Sensing}}

\begin{figure}[!t]
\begin{centering}
\includegraphics[width=0.9\columnwidth]{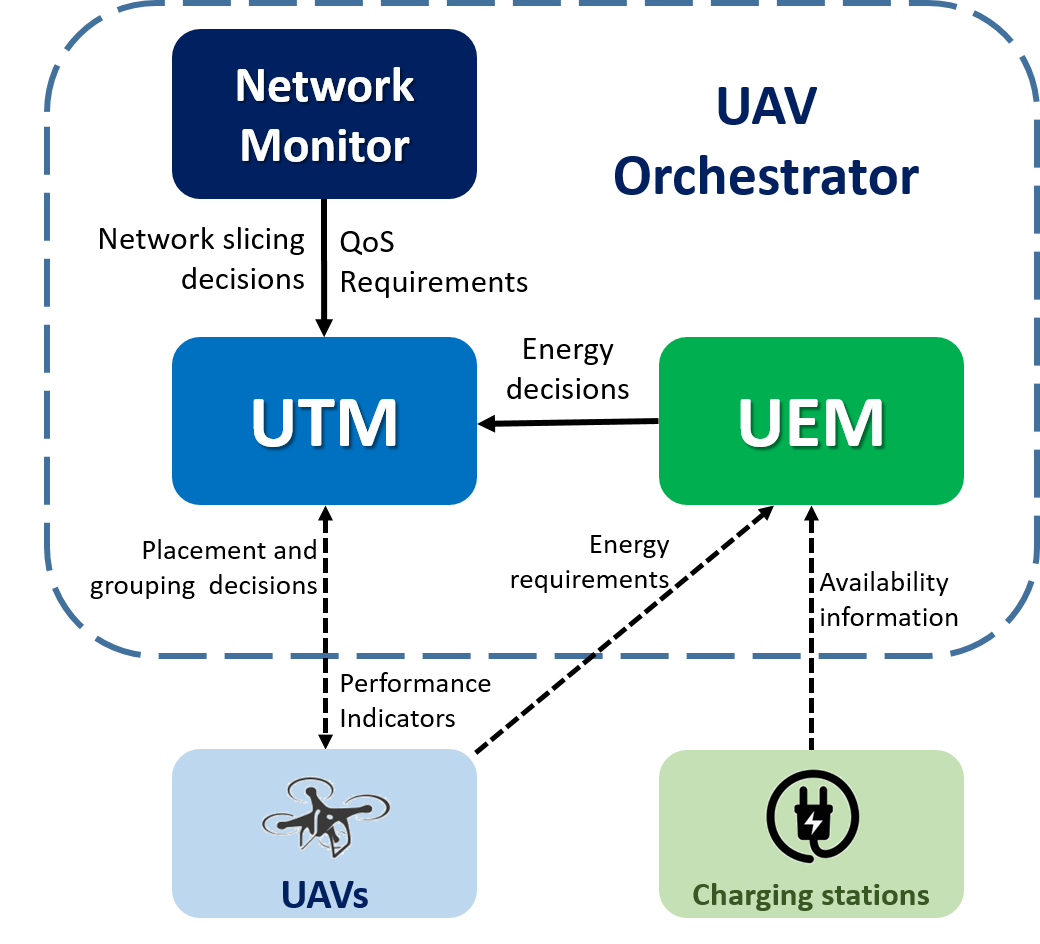}
\par\end{centering}
\caption{\protect\label{fig:UAVcontrol}UAV orchestrator components and information
flow.}
\end{figure}

As a UAV fleet scales horizontally, efficient management methods are
required to ensure its safety and high performance in terms of placement
and energy requirements~\cite{san2018uav}. Therefore, controlling
a UAV swarm is a complex operation that requires a fully automated
and omniscient orchestrator, which is a crucial software component
of any UAV network to guarantee its continuous operation.

In a nutshell, the UAV orchestrator consists of three main components:
\begin{itemize}
\item UAV Traffic Manager (UTM): The UTM is the main brain of
the orchestrator. It is responsible for the placement of the UAVs
while taking into account various restrictions and feedback from other
components, such as the UAV energy requirements, the client QoS requirements,
the weather conditions, and the airspace authorization. Moreover,
it is responsible for grouping the UAV network to allow the division
of the network for different use cases.
\item UAV Energy Manager (UEM): One of the major limitations in the UAV
operation is their energy needs. The UEM is responsible for preventing
power outages of UAVs and handling their recharging through charging
stations \cite{CS}. The decisions made by the UEM are delivered at the UTM
for appropriate mobility actions and resource allocation.
\item Network Monitor: The network monitor is aware of the QoS needs of each specific
client, as well as the needs of the different use cases that would
enable possible UAV network slicing procedures. Therefore, it informs
the UTM to take appropriate actions on the UAV placement
and grouping, respectively.
\end{itemize}
In Fig.~\ref{fig:UAVcontrol}, we present the information flow among
the different UAV orchestrator components.

\subsection{Non-programmable Metasurfaces \protect\label{subsec:NonProgrammableRIS}}

\begin{figure}[tbh]
\begin{centering}
\includegraphics[viewport=0bp 0bp 1038bp 453bp,clip,width=1\columnwidth]{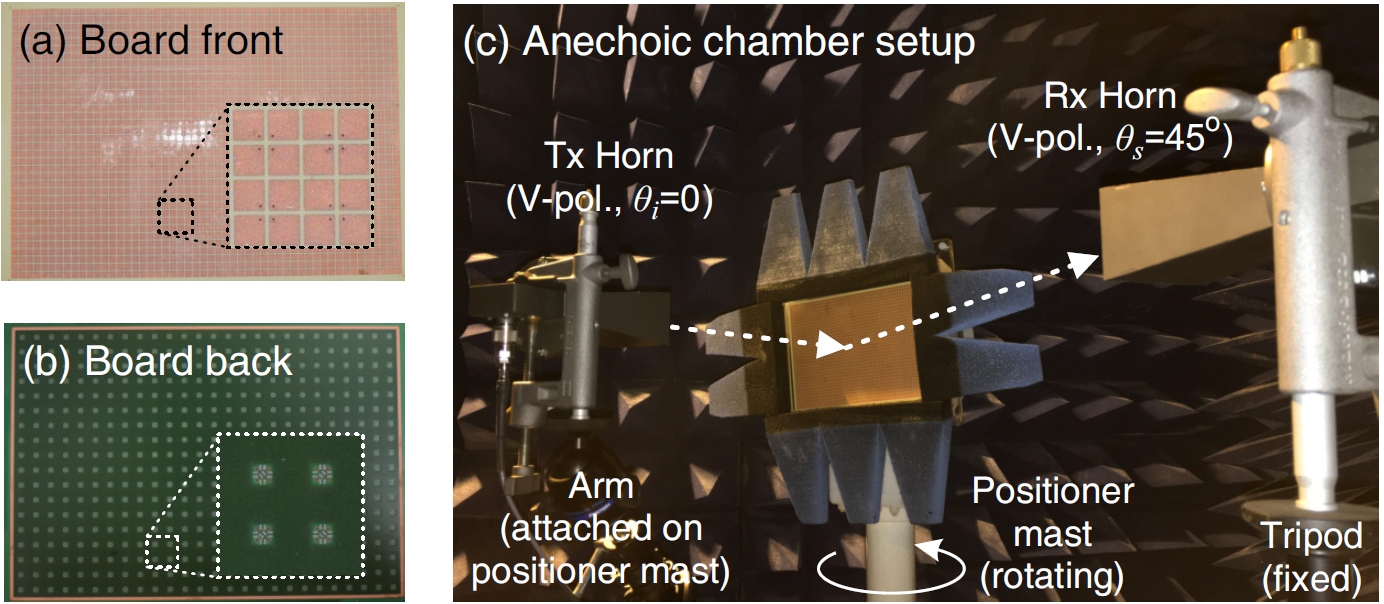}
\par\end{centering}
\caption{\protect\label{fig:PCb} (a) Front and (b) back
side of metasurface units; only the effective aperture is shown 1826
cells of 9 mm width, while the insets depict a detail of a 2-by-2
cell cluster. (c) Perspective-view of an anechoic chamber,
to acquire a measurement of the board static functionality.}
\end{figure}

%Metasurfaces are the result of Physics' attempt to build materials with engineered characteristics. Specifically, they are active enforcers of the Huygens principles meaning that they can manipulate impinging waves through a collection of embedded components (e.g. PIN diodes) to match a surface distribution that provides a needed reflected field. Metasurfaces are also capable to control every impinging wave attribute, such as amplitude, the direction of reflection, polarization, and phase. The efficiency degree with which this control is exercised is determined by the density and efficiency of the embedded components. Given the complexities of the electromagnetic field propagation reversal process, simplified ways of operation have been proposed, such as antenna arrays with phase and amplitude control per element or equivalent electric circuits. Each reduced model serves a distinct purpose in defining some of the accessible metasurface functions ~\cite{MSSurveyAllFunctionsAndTypes}.

Metasurfaces are composite materials exhibiting engineered macroscopic electromagnetic (EM) properties. They comprise specifically EM-tuned and spatially-arranged unit cells (or meta atoms) that collectively interact with the impinging wavefront to provide a given scattered (e.g., reflected) wavefront; embedding reconfigurable components (e.g., PIN diodes) in the cells allows active control of the response. Metasurfaces can control most aspects of the scattered wavefront, e.g., its amplitude, direction/phase, polarization, even frequency. The efficiency of this control scheme is determined by the unit cell architecture and the performance of the embedded components. Given the complexities of the underlying physical mechanisms, simplified methods for calculating the metasurface configuration-response relation have been proposed, such as antenna arrays with phase and amplitude control per element or equivalent circuit models. These serve distinct purposes in defining the accessible metasurface functions ~\cite{MSSurveyAllFunctionsAndTypes}.

Notably, this description refers to reprogrammable metasurface units, where
the embedded switches are controlled via software. Thus, the same
metasurface can be tuned in real-time to serve different types of wave manipulation
functionalities. However, one can replace the dynamically switching
elements with passive and non-tunable electronic packages, offering
fixed impedance to the traversing signals. When properly selected
in this manner, one can manufacture non-programmable metasurface units that
serve a very specific functionality, such as full wave absorption
from a given angle of incidence, anomalous steering for a given set
of specific angles (arrival and departure), anomalous $n$-way splitting,
etc. Such a metasurface example, manufactured with printed circuit board
technology, is given in Fig.~\ref{fig:PCb}~\cite{Pitilakis2022PRApp}.
Moreover, metasurfaces can offer the benefits of requiring no power to operate and
no need for networking capabilities. As such, if one possesses a UAV
control system and a set of metasurfaces, one can also deploy a
PWE dynamically.

\subsection{PWE deployment and configuration abstracted\protect\label{sec:PWEconfig}}

Deploying a PWE dynamically requires a macroscopic approach in its
modeling and configuration. To this end, we revisit the PWE graph,
$G\left(V,E\right)$, introduced in~\cite{liaskos2019network}. This
model considers a graph where each user device and each RIS is a vertex
$v\in V$. Moreover, we add one edge $e\in E$ between two vertexes
that are within LOS. Given this graph, it should be also considered a user $u$ that employs
$E_{u}\subset E$ edges to connect to RIS around it. Furthermore, we assume
an access point that is connected to $E_{a}\subset E$ of the total
RIS units. The PWE configuration is then equivalent to finding vertex-disjoint
paths that connect each $e\in E_{u}$ to any $e^{*}\in E_{a}$.
Once such paths are found, e.g., via the algorithms presented in~\cite{liaskos2019network},
they can be instantiated by deploying the corresponding wave steering
commands to each RIS represented by the vertexes upon each path.

However, in the studied UAV-driven PWE case, metasurface units are not fixed
in terms of location. In addition, the considered units can
serve only one specific wave manipulation functionality, such as reflecting
a given direction of wave arrival to a given direction of wave departure.
This approach makes the PWE configuration more dynamic, opening a
future research challenge. In the context of this short paper, we outline
a greedy solution as follows. We mark a set of possible UAV points
in space and add them to the graph. Then, we execute the outlined
PWE configuration algorithm. Once the paths are found, we observe how
many of the returned vertexed paths can be served by the available
set of metasurface units and their static wave steering angles. We
mark those units and corresponding vertexes as ``busy'', and
we remove the non-busy path vertexes from the graph. Subsequently,
we attempt to reconnect gaps in the intended air links, but iteratively
execute a path-finding algorithm, e.g., Dijkstra, between the gap
end-points, and then repeat the process until the path is formed (or until
a given set of retries expires).

Notably, the proposed algorithm does not account for limitations in
the UAV swarm or runtime concerns. Focusing here on a
proof of concept, we define these concerns to be in the scope of extensions
of the present study.

\section{Evaluation\protect\label{sec:Evaluation}}

We proceed to evaluate the use of a UAV-deployed PWE in a simple setting,
using a PWE simulator~\cite{liaskos2019network}. The objective is
to perform an initial comparison in a ray-tracing setting, and in
a simple geometry, using regular propagation (non-PWE) as a baseline. 

\begin{table}[t]
\centering{}\caption{\protect\label{tab:Floorplan}User location, beam orientation and
possible UAV locations. (CS origin: lower left corner).}
\begin{tabular}[b]{ccc}
\begin{tabular*}{3.8cm}{@{\extracolsep{\fill}}@{}l}
\multicolumn{1}{l}{{\small\textsc{User: Position, $\alpha$,$\phi$,$\theta$}}}\tabularnewline
\midrule 
{\small 0: {[}2.5, 7.5, 0.5{]}, 40$^{o}$, 0$^{o}$, 0$^{o}$}\tabularnewline
{\small 1: {[}7.5, 7.5, 0.5{]}, 40$^{o}$, 0$^{o}$, 180$^{o}$}\tabularnewline
\midrule 
\multicolumn{1}{l}{{\small\textsc{Comm. Pair Objective:}}}\tabularnewline
\midrule 
{\small\textcolor{black}{0$\to$1}}{\small{} : $\text{\textsc{Max. Received Power}}$}\tabularnewline
\bottomrule
\end{tabular*} & %
\begin{tabular}{c}
\tabularnewline
\includegraphics[width=0.25\columnwidth]{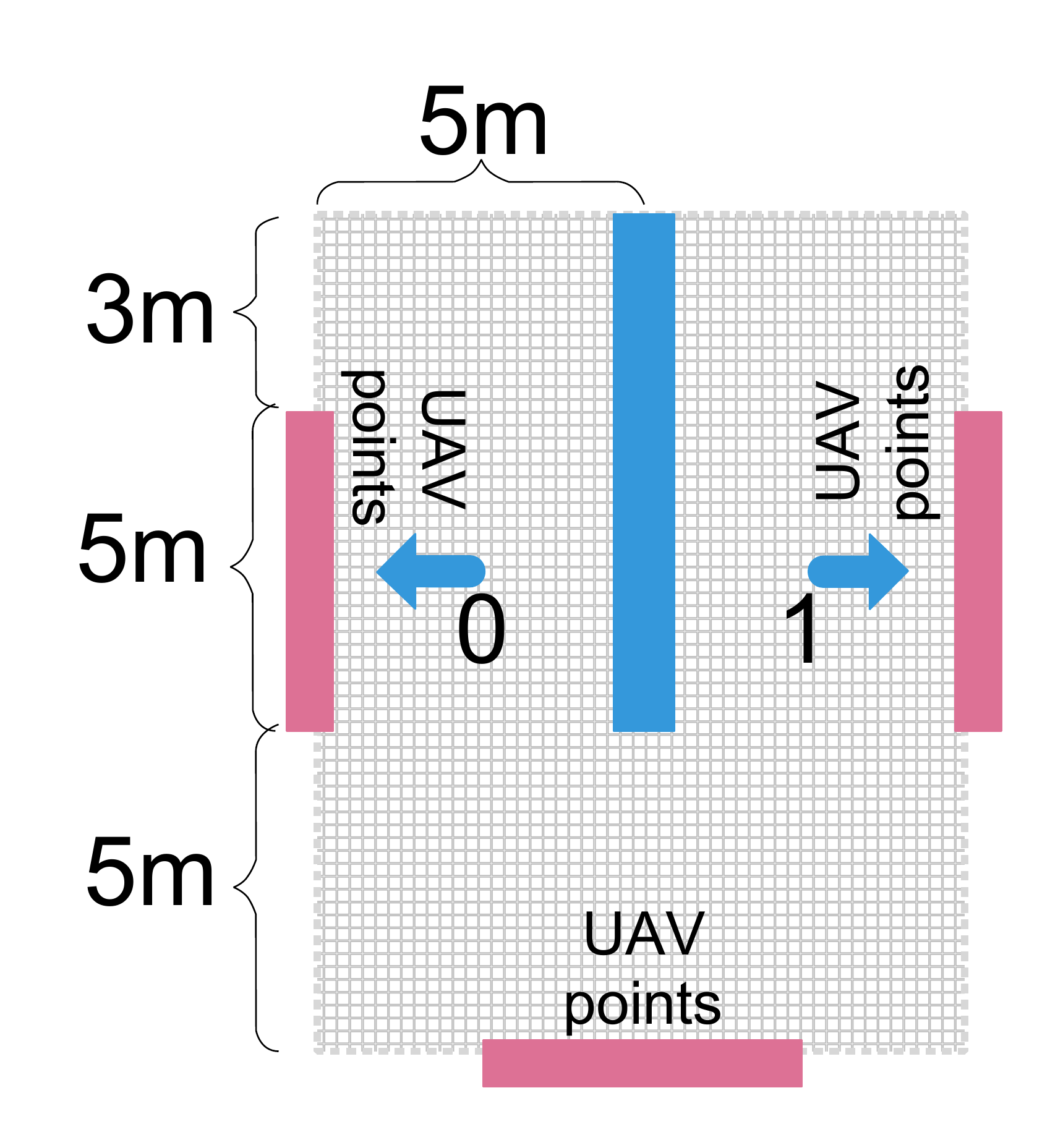}\tabularnewline
\end{tabular} & %
\begin{tabular}{c}
\tabularnewline
\hspace{-15bp}\includegraphics[width=0.2\columnwidth]{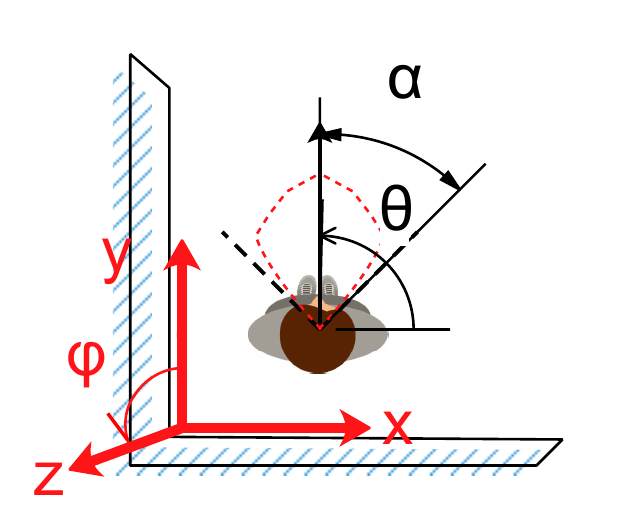}\tabularnewline
\end{tabular}\tabularnewline
\end{tabular}
\end{table}
\begin{table}[t]
\centering{}\caption{\textsc{\protect\label{tab:TSimParams}Simulation parameters.}}
\begin{tabular}{|c|c|}
\hline 
UAV flight Height & $1$~$m$\tabularnewline
\hline 
RIS Dimensions & \textbf{$1\times1\,m$}\tabularnewline
\hline 
RIS Functions & $\text{\textsc{Steer}\ensuremath{\left(From^{o},To^{o}\right)}}$\tabularnewline
\hline 
Frequency & $2.4\,GHz$\tabularnewline
\hline 
Tx Power & $-30\,dBm$\tabularnewline
\hline 
\multirow{2}{*}{Antenna type} & Single $a^{o}$-lobe sinusoid,\tabularnewline
 & pointing at $\phi^{o},\theta^{o}$ (cf. Table~\ref{tab:Floorplan})\tabularnewline
\hline 
Power loss per bounce & $1$~\%\tabularnewline
\hline 
\end{tabular}
\end{table}
We consider a transmitter (Tx)-receiver (Rx) pair in the floorplan and setup shown in Table~\ref{tab:Floorplan}, which consists of a Tx and an Rx on the plane, separated
by a perfect absorber (red wall) between them. The device positions
and setup geometry are given in Table~\ref{tab:Floorplan}. The antenna
radiation patterns are simple sinusoids of angle $\alpha$, pointed
towards opposite directions, making natural communication unlikely.
Transmission parameters are given in Table~\ref{tab:TSimParams}.

We define $15$ possible UAV locations (red areas in Table~\ref{tab:Floorplan},
with $1~m$ inter-spacing). Moreover, we assume that 15 UAVs are available,
as well as a set of static metasurface units, which are enough to serve any
possible set of wave propagation angles that can be formed in the
setup. Subsequently, we execute the process described in Section~\ref{sec:PWEconfig}.

\begin{figure}[t]
\begin{centering}
\includegraphics[width=1\columnwidth]{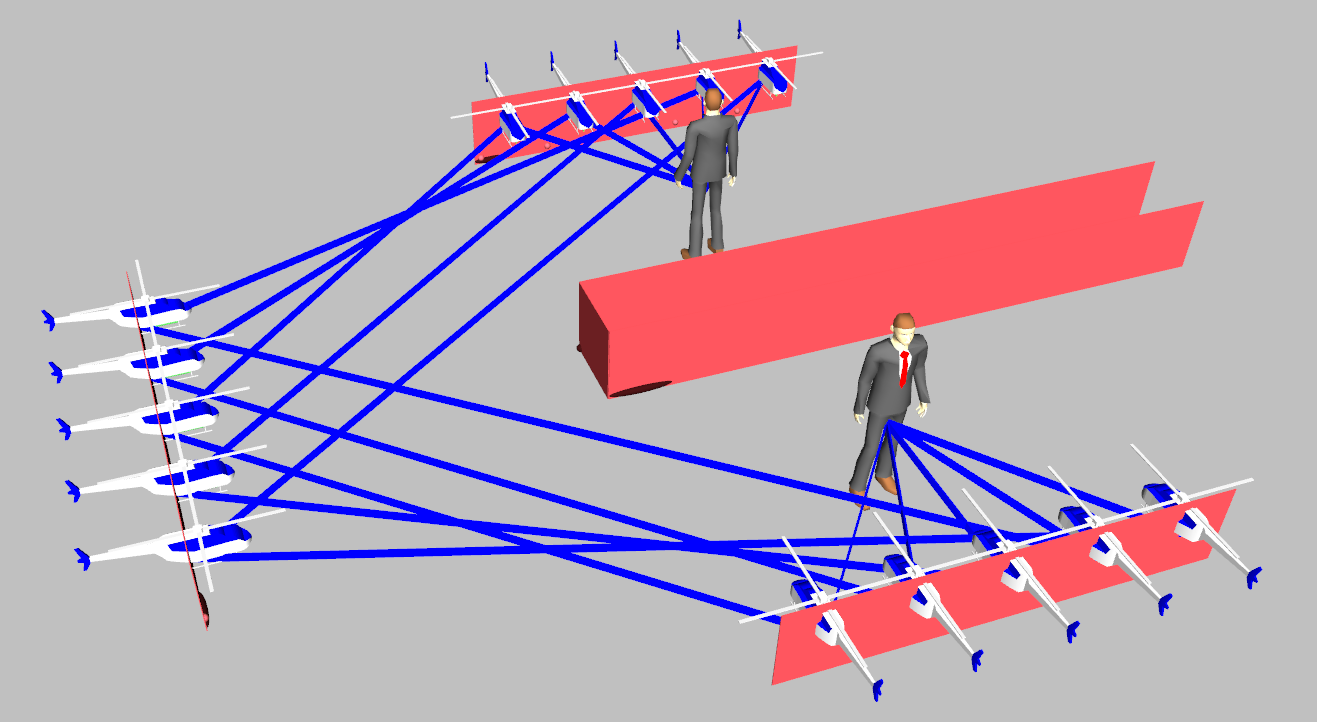}
\par\end{centering}
\caption{\protect\label{fig:uavconfig}Propagation via the proposed UAV-driven
PWE.}
\end{figure}
Visualizations of the ray-traced outcome of the UAV-driven PWE are
given in Fig.~\ref{fig:uavconfig}. Using all available UAV positions,
the dynamically formed PWE steers all major rays propagated by the Tx
to the Rx. In this way, the middle obstacle (red wall) is bypassed
during the wireless propagation phenomenon. 

\begin{table}[t]
\centering{}\caption{\textsc{\protect\label{tab:RPow}Received Power per Approach.}}
\begin{tabular}{|c|c|}
\hline 
Regular propagation (no PWE) & No signal\tabularnewline
\hline 
UAV-driven PWE & $-49.87$~$dBmW$\tabularnewline
\hline 
\end{tabular}
\end{table}
The received power per scheme is given in Table~\ref{tab:RPow}.
Notably, the employed setup offers no natural connectivity. On the
other hand, the UAV-driven PWE approach restores connectivity and
yields a received power of $-49.87\,dBmW$.

\section{Conclusion\protect\label{sec:Conclusion}}

This work presented a way of dynamically deploying PWEs with the aid of UAVs but under the condition of employing static metasurfaces which serve a specific wave manipulation
functionality. Each UAV of the swarm selects
and mounts an available passive metasurface, and proceeds to position
itself within a space to form cascaded wireless
channels for user serving. The work concluded by presenting a simulated
example of the proposed approach operation, showing that the UAV-driven PWE can effectively restore connectivity.
\section*{Acknowledgment}

This work was funded by projects HERMES (EU Horizon 2020, GA 891515), WISAR (Foundation for Research and
Technology\textendash Synergy Grants 2022) and COLLABS (EU Horizon
2020, GA 871518). 

\bibliographystyle{IEEEtran}
\bibliography{this}

\end{document}